\documentclass[final,1p,12pt]{elsarticle}

\usepackage[utf8]{inputenc}
\usepackage[T1]{fontenc}
\usepackage[english]{babel}
\usepackage{textcomp}
\usepackage{latexsym, amsmath, amssymb}
\usepackage{booktabs,multirow,multicol}
\usepackage{natbib}
\usepackage[version=4]{mhchem}

\usepackage{graphicx}
\usepackage{color,xcolor}
\usepackage{float}
\restylefloat{table}

\usepackage{mathtools}
\usepackage{exscale, relsize}
\usepackage[retainorgcmds]{IEEEtrantools}

\usepackage{array} 
\usepackage{paralist} 
\usepackage{verbatim} 
\usepackage{subcaption} 
\usepackage{rotating}
\usepackage{ulem}
\usepackage{bigstrut}
\usepackage[space]{grffile}
\usepackage{tikz}
\usetikzlibrary{shapes.geometric, arrows}
\usetikzlibrary{calc,patterns,angles,quotes,shapes,positioning}
\usepackage{xspace}
\usepackage{enumitem}

\usepackage{siunitx}
\DeclareSIUnit\atm{atm}

\usepackage{blindtext}
\usepackage{subfiles}
\usepackage{breakcites}
\usepackage[hyphens]{url}
\usepackage{amsmath}
\biboptions{sort&compress, square, comma}
\usepackage[breaklinks=true, linkcolor=blue, citecolor=blue, colorlinks=true]{hyperref}
\begin{document}

\begin{frontmatter}

\title{Computational study of the effects of density, fuel content, and moisture content on smoldering propagation of cellulose and hemicellulose mixtures}

\author[osu]{Tejas Chandrashekhar Mulky}
\author[osu]{Kyle E.~Niemeyer\corref{cor1}}
\ead{kyle.niemeyer@oregonstate.edu}

\address[osu]{School of Mechanical, Industrial, and Manufacturing Engineering, \\Oregon State University, Corvallis, OR 97331, USA}
\cortext[cor1]{Corresponding author}

\begin{abstract} 
Smoldering combustion plays an important role in forest and wildland fires.
Fires from smoldering combustion can last for long periods of time, emit more pollutants, and be difficult to extinguish.
This makes the study of smoldering in woody fuels and forest duff important.
Cellulose, hemicellulose, and lignin are the major constituents in these type of fuels, in different proportions for different fuels.
In this paper, we developed a 1-D model using the open-source software Gpyro to study the smoldering combustion of cellulose and hemicellulose mixtures. 
We first validated our simulations against experimentally obtained values of propagation speed for mixtures with fuel compositions including 100\%, 75\%, 50\%, and 25\% cellulose, with the remaining proportion of hemicellulose. 
Then, we studied the effects of varying fuel composition, density, and moisture content on smoldering combustion.
We find that propagation speed of smoldering increased with decreases in density and increases in hemicellulose content, which we attribute to the role of oxygen diffusion.
Propagation speed increased with moisture content for pure cellulose up to a certain limiting value, after which the propagation speed dropped by up to 70\%.
The mean peak temperature of smoldering increased with increases in hemicellulose content and density, and decreased with increasing moisture content.
\end{abstract}


\end{frontmatter}

\ifdefined \wordcount
\clearpage
\fi

\section{Introduction}
\label{Introduction}
Smoldering is a flameless, slow, and low-temperature form of combustion. 
It is considered to be a major fire hazard, because compared with flaming combustion it can persist for long periods of time and is difficult to suppress~\cite{Rein2016}.
The wildfires in Rothiemurchus, Scotland, that occurred during July 2006 exemplify these characteristics: the flaming part of the fire was extinguished within three days while smoldering lasted for more than 40 days---even through rain~\cite{Rein2009}.
Smoldering combustion also produces large amounts of greenhouse gases since it operates at lower temperatures resulting in incomplete oxidation.
In 1997, Indonesia's forest fires contributed around 13--40\% of the total greenhouse gases emitted by fossil fuels that year~\cite{page2002}.

Smoldering combustion can self-sustain in fuels that form char when heated since char oxidation is the main source of heat for smoldering combustion in many cases~\cite{Ohlemiller2002,Rein2016,Drysdale2011}. 
This makes study of smoldering combustion important in fuels like peat, forest duff, and woody fuels because of their abundant presence in forest. 
Such type of fuels are primarily made up of cellulose, hemicellulose, and lignin in varying proportions~\cite{Anca-Couce2012,Huang2016,Miyanshi2001}, where each constituent plays a role in the pyrolysis and combustion process~\cite{Anca-Couce2012,Huang2016}.
Smoldering combustion is generally represented by pyrolysis and oxidation reactions~\cite{Rein2016}.
Some studies have looked into the contributions of these constituents to pyrolysis.
Gani~\cite{Gani2007} found that samples with more cellulose content pyrolyzed faster than samples with more lignin.
However, we are unaware of any studies that examined how changes in fuel composition affect smoldering combustion.
These fuels also have different amounts of moisture content (MC), depending upon the porosity of the fuel and weather conditions. 
Peat, for example, can have MC ranging from 10 to 300\% depending on the weather conditions in a given region~\cite{Huang2015}.
Huang and Rein \cite{Huang2017} recently showed that downward propagation speed of smoldering increases with increasing moisture content for peat both experimentally and computationally for a range of moisture content from 0 to 70\%. 
They attributed this increase in spread rate to enhancement of thermal conductivity and reduction in the density of fuel due to addition of water~\cite{Huang2017}.

In this paper, we study smoldering combustion in cellulose and hemicellulose mixtures.
We developed a one-dimensional computational model for a reactive, porous medium with the open-source software Gpyro.
We first validate the model against experimental values of propagation speed and mean peak temperature.
Then, we look at how changes in fuel composition, density, and moisture content affect the smoldering propagation speed and mean peak temperature.
Wildlands and forests have abundant duff and woody fuels with varying fuel composition, fuel density, and moisture content. 
Understanding how these properties affect smoldering characteristics will help improve understanding of smoldering in wildland\slash forest fires and help inform large-scale models used---and decisions made---by land managers.

\section{Computational model}
In this article, we investigate the downward propagation of smoldering. 
Hence, we developed a one-dimensional computational model with a computational domain of \SI{0.0875}{\meter}. 
This domain size was chosen to match that of the experiment against which we will validate our model, where fuel was loaded in a container with the dimensions $0.2\times 0.2 \times \SI{0.0875}{\meter^3}$; Cowan et al.~\cite{Cowan2017} provide additional details on the experimental configuration. Additional information about the experiment is provided in the supplementary material. 
The top surface was open to the atmosphere while the bottom surface was insulated. 

In this model, the condensed phase and gas phase are assumed to be in thermal equilibrium (i.e., they have the same temperature). 
(Not making this assumption changes the calculated propagation speeds within 5.6\%, but at a greater computational expense.)
The shrinkage of the sample during the smoldering process is taken into consideration by decreasing cell heights ($\Delta z$)~\cite{Lautenberger2009}.
The Schmidt number is taken as unity.
All the simulations were run with an initial time-step size of \SI{0.02}{\second} and uniform cell size of \SI{e-4}{\meter}. 
These values were selected after performing a grid refinement study by reducing the spatial and initial time step by a factor of two, which only changed propagation speeds by 1.23\%. 
We provide a more detailed grid convergence study in the supplementary material.

\subsection{Governing equations}
We developed the one-dimensional model using the open-source software Gpyro~\cite{Lautenberger2009,gpyro:0.7}, which 
solves 1D transient conservation equations for condensed-phase mass Eq.~\eqref{condensed-phase mass}, gas-phase mass Eq.~\eqref{gas-phase mass}, condensed-phase species Eq.~\eqref{condensed-phase species}, gas-phase species Eq.~\eqref{gas-phase species}, condensed-phase energy Eq.~\eqref{condensed-phase energy}, and gas-phase momentum Eq.~\eqref{gas-phase momentum}, shown below; the ideal gas law Eq.~\eqref{ideal-gas law} closes the set of equations.
Gpyro is also capable of doing 2D and 3D simulations.
Lautenberger and Fernandez-Pello~\cite{Lautenberger2009} provide more details about these governing equations and how Gpyro solves them numerically.
{\allowdisplaybreaks \begin{IEEEeqnarray}{rCl}
\frac{\partial\overline{\rho}}{\partial t} &=& - \dot{\omega}^{\prime\prime\prime}_{fg} \;, \label{condensed-phase mass} \\
\frac{\partial(\rho_g\overline{\psi})}{\partial t} + \frac{\partial \dot{m}^{\prime\prime}}{\partial{z}} &=& \dot{\omega}^{\prime\prime\prime}_{fg} \;, \label{gas-phase mass} \\
\frac{\partial(\overline{\rho}Y_i)}{\partial t} &=&  \dot{\omega}^{\prime\prime\prime}_{fi} - \dot{\omega}^{\prime\prime\prime}_{di} \;, \label{condensed-phase species} \\
\frac{\partial (\rho_g\overline{\psi}Y_j)}{\partial t} + \frac{\partial{(\dot{m}^{\prime\prime} Y_j)}}{\partial z} &=& -\frac{\partial}{\partial z}(\overline{\psi}\rho_gD\frac{\partial Y_j}{\partial z})+ \dot{\omega}^{\prime\prime\prime}_{fj} - \dot{\omega}^{\prime\prime\prime}_{dj} \;, \label{gas-phase species} \\
\frac{\partial(\overline{\rho}\overline{h})}{\partial t} &=&
\frac{\partial}{\partial z}(\overline{k}\frac{\partial T}{\partial z})- \dot{Q}^{\prime\prime\prime}_{s-g}+ \sum_{k=1}^{K} \dot{Q}^{\prime\prime\prime}_{s,k}- \frac{\partial \dot q^{''}_r}{\partial z} \nonumber \\
&& +\: \sum_{i=1}^{M} ((\dot \omega ^{'''}_{fi}-\dot{\omega}^{\prime\prime\prime}_{di}) h_i) \;, \text{ and} \label{condensed-phase energy} \\
\dot{m}^{\prime\prime} &=& -\frac{\overline{K}}{v}\frac{\partial P}{\partial z} \;, \label{gas-phase momentum} \\
P \overline{M} &=& \rho_g R T_g \;, \label{ideal-gas law}
\end{IEEEeqnarray}}
where $\rho$ is the density, $M$ is the number of condensed-phase species; $X$ is the volume fraction; $\dot{\omega}^{\prime\prime\prime}$ is the reaction rate; $Y_j$ is the $j$th species mass fraction; $\psi$ is the porosity; $K$ is the permeability/number of reactions; $\overline{M}$ is the mean molecular mass obtained from local volume fractions of all gaseous species; $\dot{q}^{\prime\prime}_r$ is the radiative heat-flux; $\dot{Q}^{\prime\prime\prime}$ is the volumetric rate of heat release/absorption; $R$ is the universal gas constant; $D$ is the diffusion coefficient; $h$ is the enthalpy; $P$ is the pressure; subscripts $f$, $d$, $i$, $j$, $k$, $s$, and $g$ are formation, destruction, condensed-phase species index, gas-phase species index, reaction index, solid, and gas.
The overbar over $\rho$, $\psi$, $K$, $k$ indicates an averaged value weighted by condensed-phase volume fraction, while the overbar over $h$ indicates averaged value weighted by condensed-phase mass fraction.

\subsection{Boundary conditions}

The ambient pressure $( P_{\infty} )$ and temperature $( T_{\infty} )$ were set to \SI{1}{\atm} and \SI{293}{\kelvin}, respectively. The top surface was modeled as open to atmosphere.
For all simulations, pressure at the top surface ($z=\SI{0}{\meter}$) is set equal to the ambient pressure, Eq.~\eqref{bc_top_surface_5}.
The convective heat transfer coefficient at $z=\SI{0}{\meter}$ is set as $h_{c0}=1.52{\Delta T}^{1/3}\approx$ \SI{10}{\watt\per\meter^2\kelvin}, which takes into account cooling by the atmosphere at top surface~\cite{Huang2016}, and the mass transfer coefficient $(h_{m0})$ is set at 0.01~\cite{Huang2015}. 
The mass fractions of oxygen ($Y_{\infty,\ce{O2}}$) and nitrogen ($Y_{\infty,\ce{N2}}$) were fixed at 0.23 and 0.77, respectively.
The emissivity ($\epsilon$) is set at 0.95.


To validate the model using experimental results, we specified the boundary condition at the top surface ($z=\SI{0}{\meter}$) by setting the temperature as equal to the experimentally obtained values (via thermocouple) $T_{\text{exp}}(t)$  at $z=\SI{0}{\meter}$, using Eq.~\eqref{bc_top_surface_1}:
\begin{equation}
T\rvert_{z=0}(t) = T_{\text{exp}}(t) \;.
\label{bc_top_surface_1}
\end{equation}
Since the thermocouple at $z=\SI{0}{\meter}$ does not move as the fuel shrinks, we modeled our boundary condition to behave in the same way (i.e., remaining applied at $z = 0$.

For our remaining studies, we studied the effect of density and moisture content on smoldering. 
For those cases, we specified the boundary condition at top surface by applying a heat flux ($\dot{q_e}{\prime\prime}$) represented by Eq.~\eqref{bc_top_surface_2}. 
Once the sample is successfully ignited, for the rest of the simulation we applied a convective--radiative heat balance to the top surface, represented by Eq.~\eqref{bc_top_surface_3}:
\begin{align}
\left. -\overline{k}\frac{\partial T}{\partial z} \right\rvert_{z=0} &= -h_{c0}(T_{z=0}-T_{\infty})+\overline{\epsilon}\dot{q}_{e}^{\prime\prime} - \overline{\epsilon}\sigma(T_{z=0}^4-T_{\infty} ^{4})] \; \text{ and} \label{bc_top_surface_2} \\
\left. -\overline{k}\frac{\partial T}{\partial z} \right\rvert_{z=0} &= -h_{c0}(T_{z=0}-T_{\infty})-\overline{\epsilon}\sigma(T^4_{z=0}-T_{\infty} ^{4}) \;. \label{bc_top_surface_3}
\end{align}
Additional boundary conditions at the top surface include
\begin{align}
-\left. \left( \overline{\psi}\rho_g D \frac{\partial Y_j}{\partial z} \right) \right\rvert_{z=0} &= h_{m0} \left( Y_{j\infty}-Y_{j} \rvert_{z=0} \right) \label{bc_top_surface_4} \; \text{ and} \\
P\rvert_{z=0} &= P_{\infty} \;. \label{bc_top_surface_5}
\end{align}

For all cases, we modeled the bottom surface as insulated.
The convective heat transfer coefficient $(h_{cL})$  at the bottom surface was set at \SI{3}{\watt\per\meter^2\kelvin}~\cite{Huang2015}.
This takes into account the small amount of heat transfer across the insulated wall. 
The mass transfer coefficient $(h_{mL})$ and mass flux $(m''_L)$ were both set at 0. 
The additional equations used for the boundary conditions at the bottom surface are
\begin{align}
\left. -\overline{k}\frac{\partial T}{\partial z} \right\rvert_{z=L} &= -h_{cL}(T\rvert_{z=L}-T_{\infty}) \;, \label{bc_back_surface_1} \\
\left. -(\overline{\psi}\rho_gD\frac{\partial Y_j}{\partial z}) \right\rvert_{z=L} &= h_{mL}(Y_{j\infty}-Y_{j}\rvert_{z=L}) \;, \text{ and} \label{bc_back_surface_2} \\
\dot{m}^{\prime\prime}\rvert_{z=L} &= 0 \;. \label{bc_back_surface_3}
\end{align}

\subsection{Physical properties}

The kinetic model used in this work includes five condensed-phase species: cellulose, hemicellulose, alpha-char, beta-char, and ash. 
The model of Huang et al.~\cite{Huang2016} produces two types of char from the fuel: $\alpha$-char and $\beta$-char.
The $\alpha$-char is obtained from cellulose pyrolysis while $\beta$-char is obtained from oxidative degradation of cellulose, but we assumed the properties of $\alpha$-char and $\beta$-char to be the same~\cite{Huang2016}.
For validation, the bulk density of cellulose and hemicellulose were measured experimentally \cite{Cowan2017}.
The bulk density of char and ash were calculated using the relations $\rho_{\text{char}} \approx 0.25\times\rho_{\text{cellulose}}$ \cite{Huang2017} and $\rho_{\text{ash}} \approx \text{IC}/100\times10\times\rho_{\text{cellulose}}$, where IC stands for inorganic content~\cite{Huang-comm}.
The IC for cellulose and hemicellulose is taken as 0.3\% and 1.7\% respectively \cite{Moriana2014}. 
The bulk density of the mixture in this model $({\rho}_{\text{mix}})$ is calculated by taking into account the bulk density of cellulose and hemicellulose before mixing and mass fraction of those species $(Y_i)$ in the mixture: 
\begin{align} 
\rho_{\text{mix}} = \left( \frac{Y_{\text{cellulose}}}{\rho_{\text{cellulose}}}+\frac{Y_{\text{hemicellulose}}}{\rho_{\text{hemicellulose}}} \right)^{-1} \;.
\label{bulk_density_mixture}
\end{align}
Table~\ref{table1} provides other physical properties of the condensed-phase species, which includes solid density $(\rho_s)$, thermal conductivity ($k$), and heat capacity ($c_p$). 

\begin{table*}[htbp]
\centering
\caption{Thermophysical properties of condensed phase species}
\label{table1}
\begin{tabular}{@{}l c c c c c@{}}
\toprule
Species& Solid density & Thermal conductivity & Heat capacity & Source \\ 
& (\si{\kilo\gram\per\meter^3})  & (\si{\watt\per\meter\per\kelvin}) & (\si{\joule\per\kilo\gram\per\kelvin}) &\\
\midrule
Cellulose      & 1500 &  0.356 & 1674 & \cite{CTT}\\
Hemicellulose & 1365 & 0.34 & 1200 & \cite{aseeva2014,Thybring2014,Eitelberger2011}\\
Char      & 1300 &  0.26 & 1260 & \cite{Huang2016,bejan2003heat}\\
Ash       & 2500 &  1.2 & 880 & \cite{Huang2016,bejan2003heat}\\
\bottomrule
\end{tabular}
\end{table*}

Porosity $(\psi_i)$ and effective thermal conductivity $(k_i)$ are calculated using $\psi_i = 1-(\rho_i / \rho_{s,i})$ and $k_i=k_{s,i}(1-\psi_i)+\gamma_i\sigma T^3$, respectively, where $\gamma$ is the parameter controlling the radiation heat transfer across pores \cite{Huang2016,Lautenberger2009,Yu2006}. 
The pore diameter ($d_p$), permeability ($K$), and the parameter controlling the radiation heat transfer across pores ($\gamma$) are calculated using equations \eqref{pore size}, \eqref{permeablity}, and \eqref{gamma}, respectively, which were obtained from \cite{Huang2016,Yu2006,punmia2005soil}:
\begin{align}
d_{p,i} &= \frac{1}{S_i\times\rho} \label{pore size} \\
K &\approx \num{e-3} \times d_{p,i}^2 \label{permeablity} \\
\gamma_i &\approx 3\times d_{p,i} \label{gamma} \;,
\end{align}
where $S$ is the particle surface area.
The particle surface areas of fuel and char are assumed to be the same \cite{Huang2015}. 
The values of particle surface area of cellulose, ash obtained from cellulose, hemicellulose, and ash obtained from hemicellulose are \SIlist{0.024;0.096;0.0678;0.2712}{\meter^2\per\gram}, respectively~\cite{Sigma-Aldrich,Huang2016,S_ash}.
For simulations with moisture content, the natural expansion process during water absorption is taken into account.
To account for this process, we applied a correlation to calculate the dry bulk density $(\rho_{dc})$ and wet bulk density $(\rho_{wc})$: $\rho_{dc}=(170+40MC)/(1+MC)$ and $\rho_{wc}=(170+40MC)=\rho_{dc}(1+MC)$, respectively~\cite{Huang2017}.
This correlation was developed for peat, which has a porosity of around 0.91, close to that of cellulose at 0.88; in contrast, the porosity of hemicellulose is around 0.53.
Thus, we only used this correlation for fuels with 100\% cellulose.


 



\subsection{Chemical kinetics}

The reaction rate is expressed using Arrhenius kinetics:
\begin{gather}
\dot\omega^{'''}_{dA_k}=Z_k\frac{(\overline{\rho}Y_{A_k}\Delta z)_{\sum}}{\Delta z} \left(\frac{\overline{\rho}Y_{A_k}\Delta z}{(\overline{\rho}Y_{A_k}\Delta z)_{\sum}} \right)^{n_k} \times \nonumber\\
\exp{\left(-\frac{E_k}{RT}\right)} g(Y_{O_2})
\label{reaction_rate}
\end{gather}
where
\begin{gather}
(\overline{\rho}Y_{A_k}\Delta z)_{\sum}=\overline{\rho}Y_{A_k}\Delta z\rvert_{t=0}+\int_{0}^{t}\dot\omega{^{'''}_{fi}}(\tau)\Delta z(\tau) d\tau \;.
\end{gather}
In Eq.~\eqref{reaction_rate}, for inert atmosphere $g(Y_{O_2})$ will be equal to one and for our set of simulations $g(Y_{O_2})$ will be equal to $(1 + Y_{O_2})^{n_{O_2,k}-1}$ for an oxidative atmosphere~\cite{Huang2016,Lautenberger2009}. 
Lautenberger and Fernandez-Pello~\cite{Lautenberger2009} provide additional detail about reaction rate evaluation. 

Global kinetic descriptions of smoldering combustion, in general, include reactions for fuel pyrolysis, fuel oxidation, and char oxidation. 
We used the kinetic model of Huang and Rein~\cite{Huang2016}. In the fuel pyrolysis reaction the fuel undergoes thermal degradation in absence of oxygen to produce $\alpha$-char and gas. In the fuel oxidation reaction, fuel in the presence of oxygen undergoes thermochemical conversion to form $\beta$-char and gas. 
Both $\alpha$-char and $\beta$-char undergo further, separate oxidation reaction to form ash and gas.
The smoldering reaction model also includes a drying step if moisture content is present.
For this study, the chemical kinetic parameters for smoldering combustion
for both cellulose and hemicellulose were obtained from Huang and Rein~\cite{Huang2016}. 
(All the reactions used in the model and associated parameters are given in the supplementary material.)
For cellulose smoldering the value of the stoichiometric coefficients ($\upsilon$) were obtained from Kashiwagi and Nambu~\cite{Kashiwagi1992}.
The stochiometric coefficient of char from hemicellulose was obtained from Moriana et al.~\cite{Moriana2014}, while stochiometric coefficients for ash were obtained by using the relation IC $=\upsilon_{\alpha,hp}\upsilon_{a,\alpha-co}=\upsilon_{a,ho}\upsilon_{a,\beta-co}$, where $a$, $hp$, $ho$, and $co$ stands for ash, hemicellulose pyrolysis, hemicellulose oxidation and char oxidation, respectively~\cite{Huang2014}.
The value for the amount of oxygen consumed, $\upsilon_{\text{O}_2,k}$ consumed is calculated using the relation $\upsilon_{\text{O}_2,k}= \Delta H/(-13.1)$ \si{\mega\joule\per\kilo\gram}~\cite{Hugget1980,Huang2015}.

\section{Results}
\label{Results}

The results were first validated by comparing propagation speed and mean peak temperature obtained from experimental measurements. Then, the effects of density, fuel composition and moisture content on propagation speed and mean peak temperature were examined.
We calculated propagation speed by taking the derivative of depth with respect to time at the depth where the peak temperature at a particular time.
Then, we determined the mean peak temperature by taking the average of the peak temperatures at those depths.

\subsection{Validation against experiments}
\begin{figure}[htb]
\centering
\includegraphics[width= 0.9\linewidth]{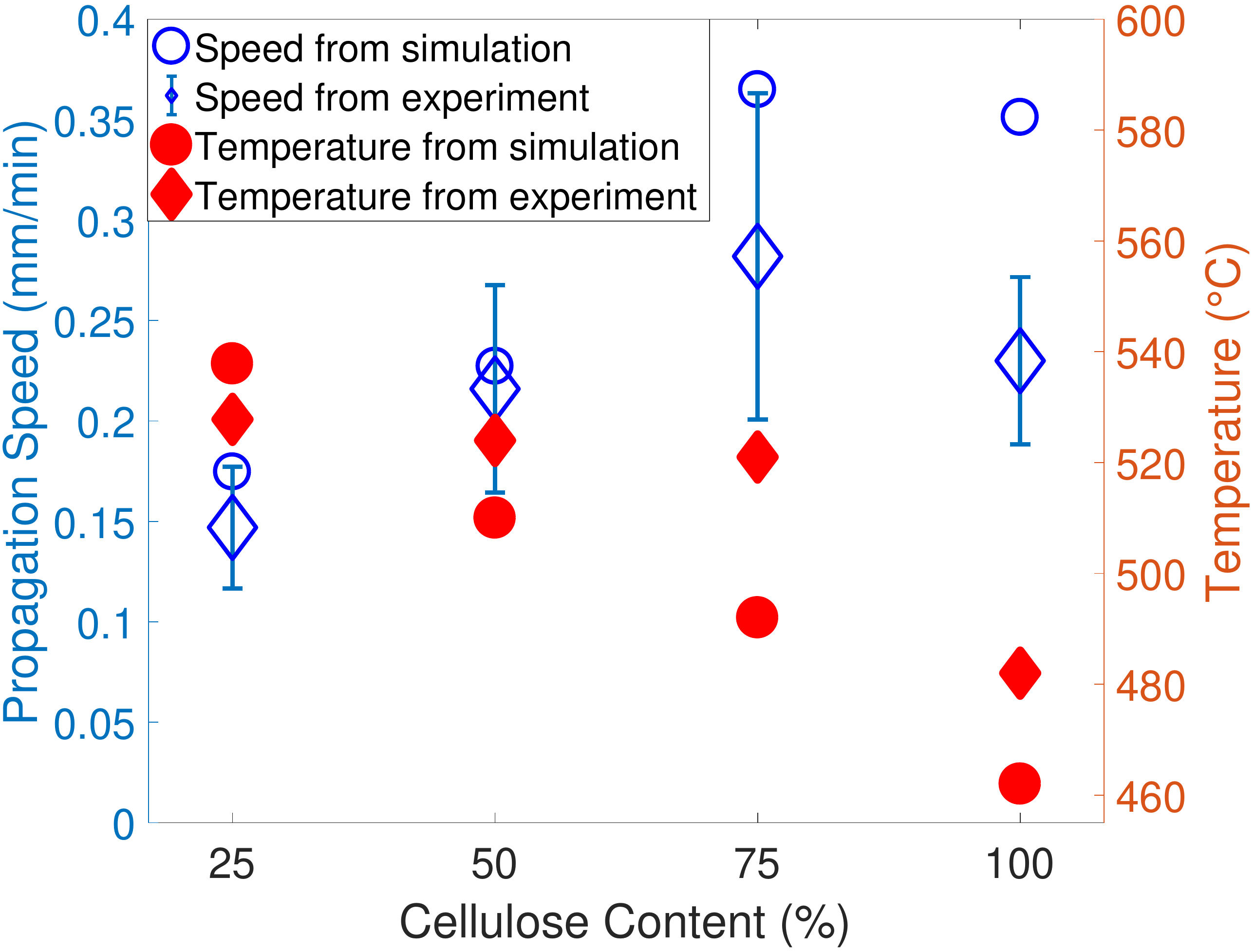} 
\caption{Comparison of mean propagation speeds and mean peak temperatures obtained from experiments and simulations.}
\label{validation}
\end{figure}

The validation against the experimental results were done by comparing the downward average propagation speed and mean peak temperature. 
In the experiments, four thermocouples were placed at \SIlist{0.0;2.5;5.0;7.5}{\centi\meter} from the top surface.
Figure~\ref{validation} shows the mean propagation speed and the mean peak temperature measured from the experiments and calculated from our simulations for a range of fuel compositions: 100\%, 75\%, 50\%, and 25\% cellulose, with the remainder as hemicellulose. 
(We did not validate for 100\% hemicellulose due to a lack of experimental data.)
From Figure~\ref{validation} we can see that for 100\% cellulose content, the model overestimates the propagation speed, but as the cellulose content drops the predicted velocities fall within the experimental error bars. 
The reason for this error could be that the cellulose samples used in the experiments are fibrous whereas the particles used to obtain the specific surface area of cellulose, char, and ash were assumed spherical~\cite{S_ash,Sigma-Aldrich}.
Note that the pore size and permeability was calculated using Eq.~\eqref{pore size} and Eq.~\eqref{permeablity}, respectively, for all the condensed-phase species.
Other reasons could be the presence of moisture and inorganic content in the fuel, which this model does not consider.
The predicted mean peak temperatures lay within 5.5\% of the experimentally measured values.

\subsection{Effect of fuel composition and density}

To examine the effects of fuel composition and density on smoldering combustion, we varied the fuel composition between 100\% and 25\% cellulose in increments of 25\%, where the remaining portion was hemicellulose, and varied the density between 200 and \SI{500}{\kg\per\m^3} in increments of \SI{100}{\kg\per\m^3}. 
To ignite the sample, we applied a heat flux of \SI{15}{\kilo\watt\per\meter^2} at the top layer for the span of 15 minutes.
\begin{figure}[htb]
\centering
\includegraphics[width= 0.9\linewidth]{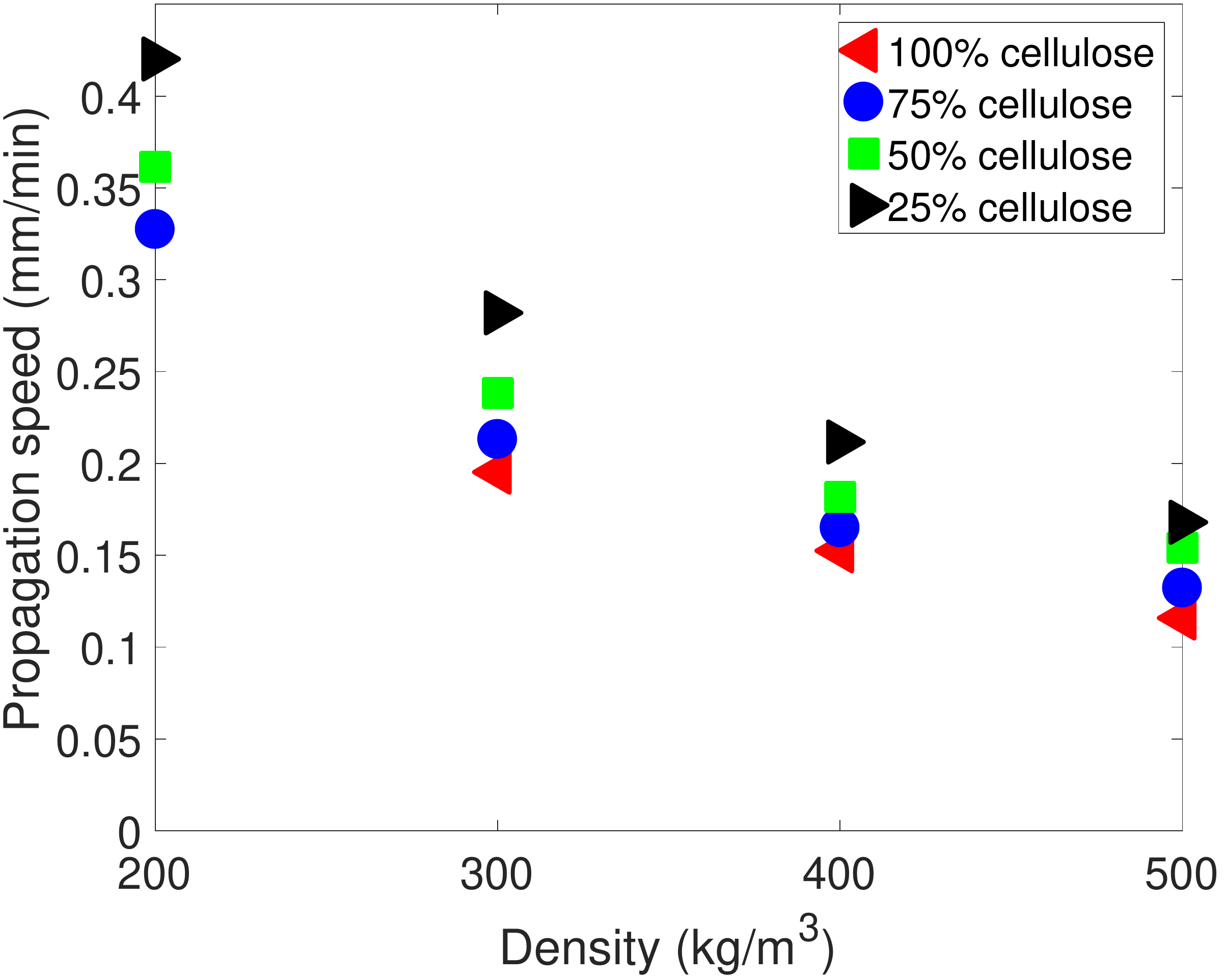}
\caption{Effect of density on mean propagation speed.}
\label{speed}
\end{figure}

\begin{figure}[htb]
\centering
\includegraphics[width= 0.9\linewidth]{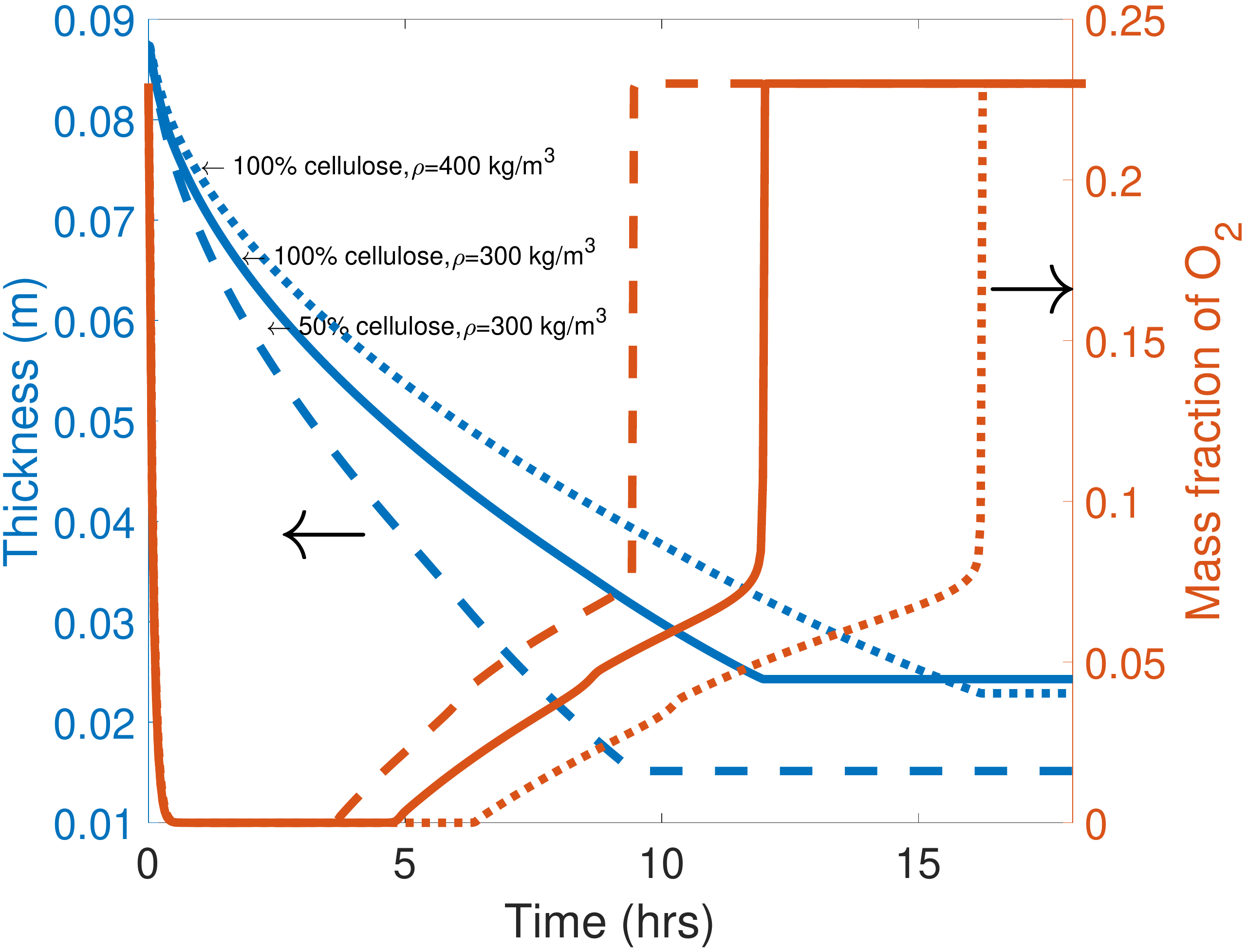}
\caption{Mass fraction of oxygen at depth 5 cm from the top and thickness over time for fuels with 100\% cellulose of density
\SI{300}{\kilo\gram\per\meter^3} and \SI{400}{\kilo\gram\per\meter^3} and 50\% cellulose of density \SI{300}{\kilo\gram\per\meter^3}.}
\label{o2_thick}
\end{figure}

Figure~\ref{speed} shows that the propagation speed drops as the density increases for all fuel compositions.
The propagation speed drops by around 60\% for all the calculated fuel compositions when the density increases from \SI{200}{\kilogram\per\meter^3} to \SI{500}{\kilogram\per\meter^3}.
For the aforementioned boundary condition, 100\% cellulose at density \SI{200}{\kilogram\per\meter^3} did not ignite.
This decrease in the propagation speed with increase in density could be due to the fact that as the density of the fuel sample increases the pore size and permeability of the fuel sample decrease, as expressed in Eqs.~\eqref{pore size} and \eqref{permeablity}.
Due to this relationship, the availability of the oxygen drops as the density increases.
Since the smoldering spread rate depends on the oxygen supply~\cite{Rein2009}, less availability of oxygen leads to a reduction in propagation speed.
Figure~\ref{o2_thick} shows the mass fraction of oxygen with respect to time at depth \SI{5}{\centi\meter} from the top surface for 100\% cellulose. 
At any point after ignition, more oxygen is available for the fuels with lower density. 

Figure~\ref{speed} shows that, for any density between 200 and \SI{500}{\kilogram\per\meter^3}, fuels with higher cellulose content have slower propagation speeds compared with fuels with higher hemicellulose content. 
Hemicellulose pyrolyzes at lower temperatures compared with cellulose~\cite{Yang2007}, and as a result there will be more mass loss at an earlier stage from pyrolysis for samples with more hemicellulose content.
This would result in more availability of oxygen for a particular depth since the sample would shrink faster, which can also be seen in Figure~\ref{o2_thick}. 
Figure~\ref{o2_thick} shows that the 50\% hemicellulose case has higher mass fraction of oxygen available at earlier times compared with 0\% hemicellulose content.
More oxygen will promote oxidation and lead to faster propagation.
\begin{figure}[htb]
\centering
\includegraphics[width= 0.9\linewidth]{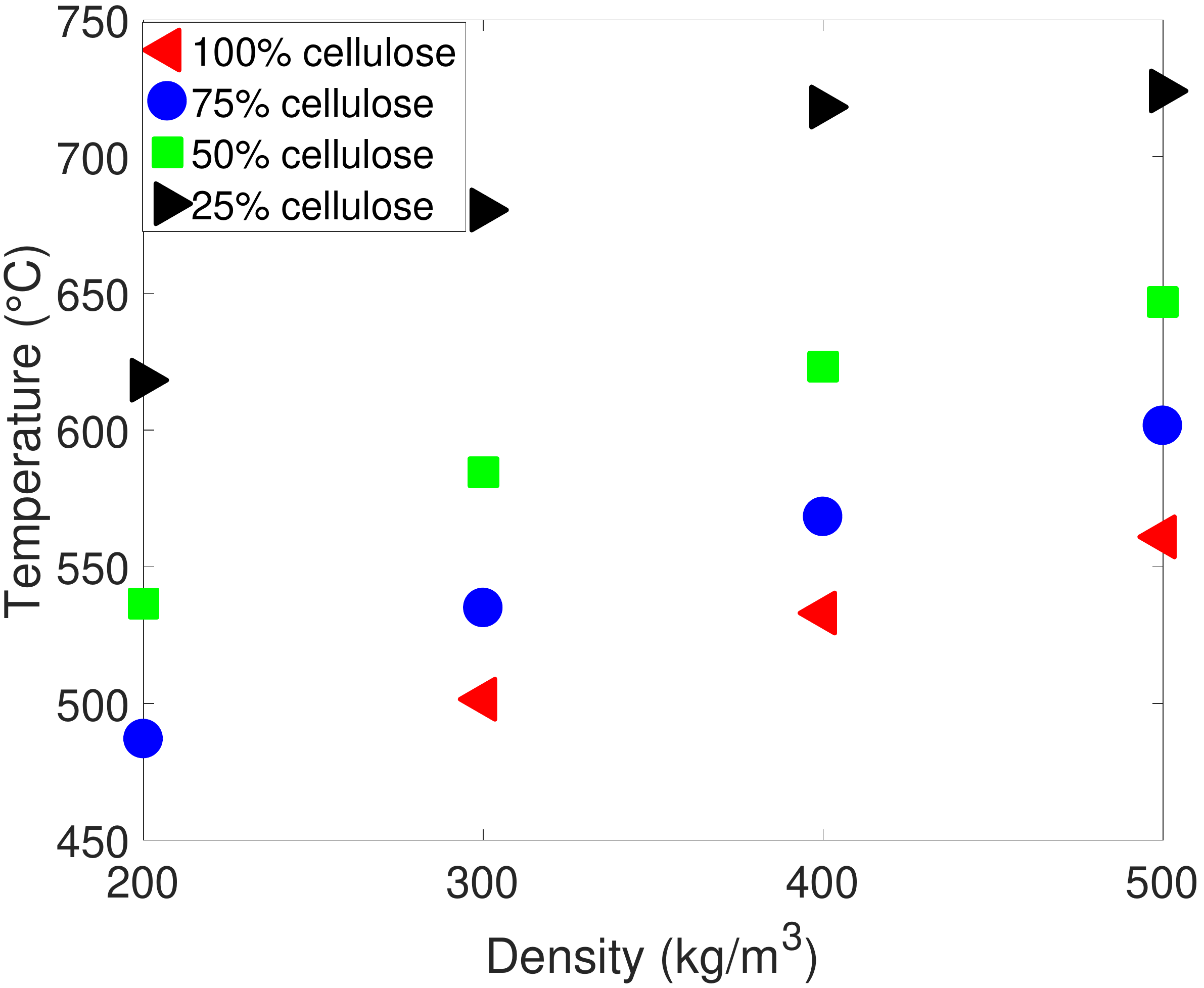}
\caption{Effect of density on mean peak temperature.}
\label{temp}
\end{figure}

Figure~\ref{temp} shows the effects of density and fuel composition on mean peak temperature.
The mean peak temperature increases with increasing density as well as hemicellulose content.
When hemicellulose content and density is increased, the amount of ash produced also increases because in general hemicellulose has more natural inorganic content than cellulose. 
Natural inorganic content is directly proportional to the amount of ash produced~\cite{Huang2014}; further, increasing density makes more fuel available, which results in formation of more ash.
Figure~\ref{o2_thick} shows that the final thickness---which is the area occupied by ash per unit length---is less for fuels with higher density and hemicellulose content, indicating that the densities of ash produced for these fuels are higher.
This higher-density ash, which is formed in the top layer, would insulate the sample and retain the heat produced from oxidation.
This would result in higher smoldering temperatures, as seen in Figure~\ref{temp}.

\subsection{Effect of moisture content}

We considered the effects of moisture content by adding a drying step to the reaction scheme.
The reaction parameters for the drying step were obtained from Huang and Rein~\cite{Huang2016}.
Adding water to the sample leads to an expansion in the fuel, which in turn decreases the density of the fuel after the water evaporates.
Here, we adopted the co-relation that takes into account this expansion from that used by Huang and Rein~\cite{Huang2017} for peat.

\begin{figure}[htb]
\centering
\includegraphics[width= 0.9\linewidth]{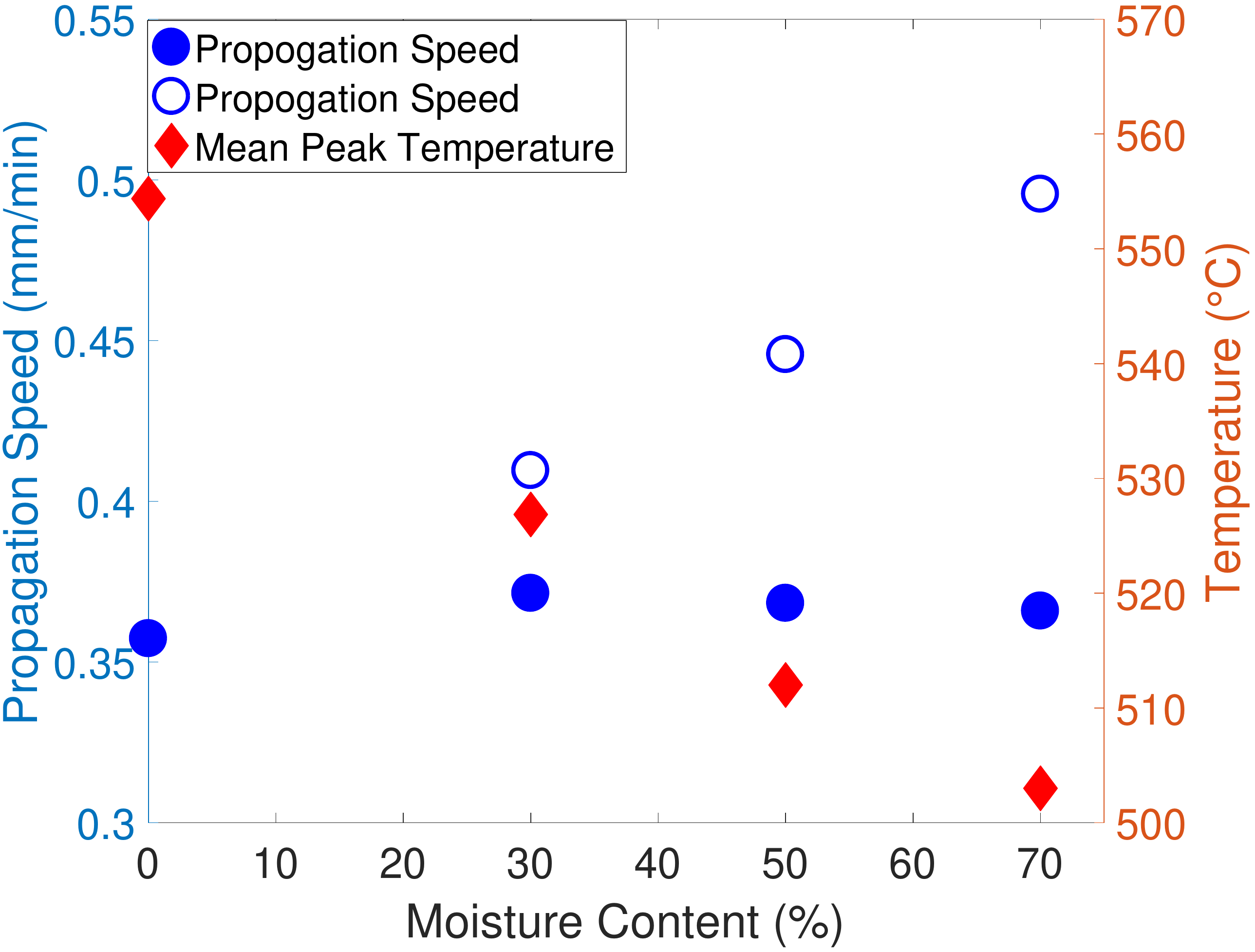}
\caption{Effect of moisture content on mean peak temperature and propagation speed. For propagation speed, filled circles indicate moisture content, while unfilled circles indicate no moisture content but with same amount of natural expansion that would occur if moisture content was added.}
\label{moisture}
\end{figure}

Figure~\ref{moisture} shows the effect of moisture content on propagation speed.
The propagation speed increases by 4\% as moisture content is increased from 0 to 30\%.
One of the reasons for this increase is that when water is mixed with the fuel, the fuel will expand, which would eventually lead to reduction in density of the fuel once the water evaporates~\cite{Huang2017}.
In addition, thermal conductivity of the wet fuel increases with moisture content due to the added water~\cite{Huang2017}.
Huang and Rein also observed increases in downward propagation speed of peat with increasing moisture content both experimentally and computationally~\cite{Huang2017}.
After 30\% moisture content, the propagation speed did not significantly change further;
the propagation speed drops by around 1.5\% when the moisture content was increased from 30\% to 70\%.
Figure~\ref{moisture} also shows how the propagation speed of 100\% cellulose changes due to reduced density due to moisture content, but without the other effects of moisture.
The difference between these two velocities indicates how the other effects of moisture content counter the effect of expansion.
As we increase the moisture content, both the effects of expansion and moisture content on smoldering grow. 
However, the increasing difference between the trends shows that the other effects of adding water---such as making the drying step more endothermic, in turn leading to a drop in the overall temperature as Figure~\ref{moisture} shows---overcome the expansion effect at higher moisture contents to reduce the propagation speed.

(For simulations with high moisture content, i.e., when moisture content is greater than the fiber-saturation point, moisture would be present as capillary water, which is not well approximated with chemical reactions.)
Temperature, on the other hand, continuously decreases as moisture content is increased from 0 to 70\%.
This is because as moisture content increases, more heat is needed to evaporate the water, which reduces the overall temperature.

\section{Conclusion}
\label{Conclusion}

In this article, we studied the downward smoldering propagation of cellulose and hemicellulose mixture.
First, we validated the model by comparing the values of propagation speed and mean peak temperature against experimentally obtained values for fuel compositions of 100, 75, 50, and 25\% of cellulose with remaining portion being hemicellulose. 
The predicted values of propagation for 75, 50, and 25\% cellulose agree with the experimental results within the measurement uncertainty. 
We suggest that the model overpredicts the values of propagation speed for 100\% cellulose due to the calculation that assumed solid particle shapes as spherical, while in reality the particles are fibrous in shape.

Next, we examined the effects of changing density, fuel composition, and moisture content on smoldering propagation speed and  mean peak temperature.
Propagation speed of smoldering combustion decreases with increases in density and cellulose content. 
The possible reason for this is lack of availability of oxygen. In the case of density, as the density increases the permeability and pore size drop, which limits the available oxygen. As hemicellulose content increases, more oxygen becomes available due the additional mass that pyrolyzes at a given time, since hemicellulose undergoes pyrolysis at lower temperatures than cellulose. 
The mean peak temperature increases with density and hemicellulose content, possibly due to more and denser formation of ash on the surface, which acts as an insulator.
In the case of moisture content on smoldering combustion for 100\% cellulose, the propagation speed increases by about 4\% as moisture content increases from 0 to 30\%.
This is caused by expansion of the fuel when water is added, which reduces the density of fuel when the water evaporates. After this point, the propagation speed only drops by about 1.4\% as moisture content increases from 30\% to 70\%, indicating a lack of sensitivity to moisture content at values above 30\%.

\section*{Acknowledgements}
This research was funded by the Strategic Environmental Research and Development Program (SERDP) award RC-2651 under contract number W912HQ-16-C-0045. 
We also thank David Blunck, Benjamin Smucker, and Daniel Cowan from Oregon State University for providing their experimental temperature measurements data for validation.

The views, opinions, and/or findings contained in this report are those of the authors and should not be construed as an official Department of Defense position of decision unless so designated by other official documentation. 

\section*{Supplementary material}

The supplementary material for this article contains a description of the experimental setup used to provide validation data, the values of kinetic parameters used in the model, and a grid convergence study.

\bibliographystyle{elsarticle-num-PROCI}
\bibliography{references.bib}


\end{document}